\documentclass{article}
\usepackage{amsmath,amsthm}
\usepackage{amssymb,latexsym}
\usepackage[mathscr]{eucal}
\usepackage{setspace}
\usepackage{graphics}
\usepackage{array}

\DeclareMathSizes{13}{13}{9}{8}

\newcommand{\pa}{\partial}

\newcommand{\dpr}{^{\prime\prime}}

\newcommand{\pr}{^\prime}
\newcommand{\rp}{\right)}
\newcommand{\lp}{\left(}
\newcommand{\rb}{\right]}
\newcommand{\lb}{\left[}
\newcommand{\rel}{\right\}}
\newcommand{\lel}{\left\{}

\newcommand{\beq}{\begin{equation}}
\newcommand{\eq}{\end{equation}}
\newcommand{\bfv}{{\bf v}}

\newcommand{\bfj}{{\bf J}}
\newcommand{\bfb}{{\bf B}}

\newcommand{\bfe}{{\bf E}}
\newcommand{\bfihat}{\hat {\bf i}}
\newcommand{\na}{\nabla}
\newcommand{\ti}{\times}

\newcommand{\bfve}{\bfv_e}
\newcommand{\bfvi}{\bfv_i}
\newcommand{\vet}{\frac{\pa\bfve}{\pa t}}
\newcommand{\vit}{\frac{\pa\bfvi}{\pa t}}
\newcommand{\pe}{p_e}
\newcommand{\pui}{p_i}
\newcommand{\lc}{\frac{1}{c}}
\newcommand{\pso}{\psi_1}
\newcommand{\pho}{\phi_1}
\newcommand{\bo}{b_1}
\newcommand{\wo}{w_1}
\newcommand{\bpso}{\Psi_1}
\newcommand{\bpho}{\Phi_1}
\newcommand{\bbo}{B_1}
\newcommand{\kt}{k^2}
\newcommand{\mut}{\mu^2}
\newcommand{\muh}{\mu^{\frac{1}{2}}}
\newcommand{\munh}{\mu^{-\frac{1}{2}}}
\newcommand{\om}{\omega}
\newcommand{\ve}{\varepsilon}
\newcommand{\st}{\sigma^2}

\usepackage{graphicx}
\graphicspath{{converted_graphics/}}
\begin{document}

\title{Hall Resistive Tearing Mode: A Variational Formulation}         
\author{Bhimsen K. Shivamoggi\footnote{\normalsize Permanent Address: University of Central Florida, Orlando, FL 32816-1364}\\
Los Alamos National Laboratory\\
Los Alamos, NM 87545}        
\date{}          
\maketitle

\large{\bf Abstract}

A unified linear tearing-mode formulation is given incorporating both resistivity and Hall effects. A variational method is used that appears to be best suited to deal with the difficulties peculiar to the {\it triple-deck} structure associated with the Hall resistive tearing mode but also to lead to a convenient analytical dispersion relation for the Hall resisitive tearing mode. 

This analytical dispersion relation - 

\begin{itemize}
  \item recovers the Furth-Killeen-Rosenbluth \cite{Fur} result for the resistive branch;
  \item gives a growth rate for the Hall branch which  appears to be consistent with the growth rate of the electron-inertia driven tearing mode given previously (Coppi \cite{Cop});
  \item recovers the scaling relation for the transition from the resisitive regime to the Hall regime numerically established by Fitzpatrick \cite{Fit} in a driven Hall resistive reconnection situation.
\end{itemize}

\pagebreak

\section{Introduction}      

Fast magnetic reconnection processes in laboratory (ex. sawtooth collapse in tokamak discharges) and space (ex: solar flares and magnetospheric substorms) can be described using collisionless plasma models (Yamada et al.\cite{Yam}, Shibata \cite{Shi} and Nishida \cite{Nis}). Collisionless plasma processes cannot be described in terms of a single-fluid formulation of resistive MHD. In a high-$\beta$ collisionless plasma, on length scales shorter than the ion skin-depth $d_i$, the electrons decouple from the ions and the electron dynamics is governed by Hall currents (Sonnerup \cite{Son}) and the characteristic mode is the whistler wave. The dispersive nature of the whistler wave leads to current-sheet broadening - this is confirmed by laboratory experiments (Urrutia et al. \cite{Urr}) and satellite observations at the magnetopause and the magnetotail plasma sheet (Sonnerup et al. \cite{Son2}, Fairfield et al. \cite{Fai}). The ensuing reconnection process is therefore faster (Mandt et al. \cite{Man} and Biskamp et al. \cite{Bis}). The reconnection rate is primarily controlled by ions (which are decoupled from the electrons) and is independent of the mechanism that breaks the frozen-in condition of the magnetic field lines (resistivity or electron inertia) - this was confirmed by numerical calculations (Birn et al. \cite{Bir}).

Terasawa \cite{Ter} considered the effect of the Hall current on the evolution of the resistive tearing mode which is believed to play an important role in reconnection processes and found that Hall effects enhance the tearing-mode growth rate. An important observation made by Terasawa \cite{Ter} in course of this investigation was that the Hall current leads to a 3D magnetic field structure although one may assume 2D for the spatial dependence of the mode structure - the \textit{quadrupolar} out-of-plane magnetic field pattern (which is a signature of whistler-controlled dynamics) was confirmed by {\it in situ} measurements in the magnetotail (Oieroset et al. \cite{Oie}) and laboratory experiments (Ren et al. \cite{Ren}). The Hall current was found to nonlinearly couple the out-of-plane magnetic (as well as plasma velocity) field to their otherwise autonomous planar field counterparts. However, the numerical procedure used by Terasawa \cite{Ter} to calculate the tearing mode growth rate did not afford reliable determination of the scaling relations associated with this growth rate. The Hall resistive tearing mode was also considered by Hassam \cite {Has} who applied, on the other hand, the Furth-Killeen-Rosenbluth procedure for this problem and his treatment is restricted by several assumptions - 
\begin{itemize}
  \item ions are cold,
  \item plasma flow in the non-ideal MHD region is negligible,
  \item constant - {$\psi$} ({$\psi$} being the stream function for the in-plane or poloidal magnetic field) approximation. 
  \end {itemize}
 Besides, it would appear that the existence of a {\it triple-deck} structure (borrowing the terminology from boundary layer theory in fluid dynamics) in Hall resistive MHD comprising
\begin{itemize}
  \item the ideal MHD region,
  \item The Hall layer,
  \item the resistive layer,
  \end{itemize}
 (another important observation made by Terasawa \cite{Ter}) poses difficulties in implementing the FKR procedure and producing a convenient analytical dispersion relation for this problem. A variational method (Hazeltine and Strauss \cite{HazS} and Hazeltine and Ross \cite{HazR}), on the other hand, appears to be best suited to deal with the difficulties peculiar to the {\it triple-deck} structure associated with the Hall resistive tearing mode. The present paper aims at addressing this and shows that application of the variational method also leads to a convenient and apparently sound analytical dispersion relation for the Hall resistive tearing mode.

\section{Governing Equations for Hall MHD}
Consider an incompressible, two-fluid, quasi-neutral plasma. The governing equations for this plasma dynamics are (in usual notation) - 

\beq
nm_e \lb \vet + (\bfve\cdot\na)\bfve\rb = -\na\pe-ne (\bfe+\lc\bfve\ti\bfb) + ne\eta \bfj
\eq

\beq
nm_i \lb \vit + (\bfvi\cdot\na)\bfvi\rb = -\na\pui+ne(\bfe+\lc\bfvi\ti\bfb)-ne\eta\bfj
\eq

\begin{align}
\na\cdot\bfve &=0\\
\na\cdot\bfvi &=0\\
\na\cdot\bfb &=0\\
\na\ti\bfb &=\lc\bfj\\
\na\ti\bfe &=-\lc\frac{\pa\bfb}{\pa t}
\end{align}

\noindent
where,
\beq
\bfj\equiv ne(\bfvi - \bfve).
\eq

Neglecting electron inertia ($m_e \rightarrow 0$), equations (1) and (2) can be combined to give an ion equation of motion - 

\beq
nm_i \lb \vit +(\bfvi\cdot\na)\bfvi\rb = -\na(p_i + p_e) +\lc \bfj\ti\bfb
\eq

\noindent
and a generalized Ohm's law - 

\beq
\bfe + \lc \bfv_{i}\ti\bfb = \eta\bfj +\frac{1}{nec}\bfj\ti\bfb.
\eq

Non-dimensionalize distance with respect to a typical length scale $a$, magnetic field with respect to a typical magnetic field strength $B_0$, time with respect to the reference Alfv\'en time $\tau_A \equiv a/V_{A_0}$ where $V_{A_0} \equiv B_0/\sqrt{\rho}$ and $\rho \equiv m_i n$, and introduce the magnetic and velocity stream functions according to

\beq
\left.
\begin{array}{l}
\bfb = \na\psi \ti\bfihat_z + b\bfihat_z\\
\bfvi = \na\phi \ti\bfihat_z + w\bfihat_z
\end{array}\rel
\eq

\noindent
and assume the physical quantities of interest have no variation along the $z$-direction. Equations (9) and (10), then yield

\beq
\frac{\pa\psi}{\pa t} + [\psi,\phi] + \sigma [b,\psi] = \varepsilon \na^2\psi
\eq

\beq
\frac{\pa b}{\pa t} + [b,\phi] + \sigma [\psi,\na^2\psi] +[\psi, w] = \varepsilon\na^2 b
\eq

\beq
\frac{\pa}{\pa t} (\na^2\phi) +[\na^2\phi,\phi] = [\na^2\psi,\psi]
\eq

\beq
\frac{\pa w}{\pa t} + [w,\phi] = [b,\psi]
\eq

\noindent
where,

\beq
\left.
\begin{array}{r}
[A,B] \equiv \na A \ti \na B\cdot\bfihat_z\\
\displaystyle{\sigma \equiv \frac{d_i}{a},\varepsilon\equiv\frac{\eta c^2 \tau_A}{a^2}}.\notag
\end{array}
\rel
\eq

\section{Hall Resistive Tearing Modes}

Let us write,

\begin{align}
\psi &= \psi_0 (x) + \pso(x)e^{\omega t} \cos y\\ \notag
\phi &= \pho(x) e^{\omega t} \sin y\\
b &= \bo (x)e^{\omega t} \sin y\notag\\ 
w &= \wo (x)e^{\omega t} \cos y .\notag
\end{align}

Equations (12) - (15) then give, on linearlizing in $\pso,\ \pho,\ \bo,$ and $\wo$,

\beq
\omega \pso - F(\pho -\sigma\bo) = \varepsilon (\pso\dpr - \pso)
\eq

\beq
\omega \bo + \sigma F(\pso\dpr - \pso)+ F\wo = \varepsilon (\bo\dpr - \bo) 
\eq

\beq
-\omega (\pho\dpr - \pho) = F(\pso\dpr - \pso)- F\dpr\pso
\eq

\beq
\omega \wo - F\bo = 0
\eq

\noindent
where primes denote differentiation with respect to $x$, and 

\beq
\psi_0' = -F.
\eq

Near the magnetic neutral surface $x=0$, where $F\approx x$, and the flow and field gradients become large, equations (17) - (20) become

\beq
\omega\pso + x\pho - \sigma x\bo = \varepsilon \pso\dpr
\eq

\beq
\omega\bo - \sigma x \psi\dpr = \varepsilon \bo\dpr
\eq

\beq
\omega\pho\dpr = x \pso\dpr
\eq

Equation~(23)~shows that $b_1$ is an odd function of $x$ (since $\psi_1$ is an even function of $x$) while (16) specifies that $b_1$ is an odd function of $y$ as well - this sets forth the \textit{quadrupolar} structure of the out-of-plane magnetic field crucial for Hall MHD.

We now use a variational method ([16] and [17]) to develop an analytical dispersion relation for the Hall resistive tearing mode. On Fourier transforming the variables, according to

\beq
Q(k) = \frac{1}{\sqrt{2\pi}} \int_{-\infty}^\infty e^{-ikx}q(x) dx
\eq

\noindent
equations (22) - (24) give

\beq
\omega \bpso +  i\bpho\pr - \sigma i \bbo\pr = -\ve \kt \bpso
\eq

\beq
\omega \bbo + i\sigma (\kt\Psi_1)\pr = -\ve\kt\bbo
\eq

\beq
\om \kt\bpho = i (\kt \bpso)\pr.
\eq

Equations (26) - (28) can be combined into the self-adjoint form - 

\beq
\lb \lp \frac{1}{\om\kt} + \frac{\st}{\om+\ve\kt}\rp J\pr\rb\pr - \lp\ve+\frac{\om}{\kt}\rp J = 0
\eq

where,

\beq
J\equiv \kt\bpso.
\eq

Equation (30) can be obtained by extremizing the bilinear functional 

\beq
S=\int_{-\infty}^{\infty} \lb \lp\frac{1}{\om\kt} + \frac{\st}{\om + \ve\kt}\rp J^{\prime^2} + \lp \ve+\frac{\om}{\kt}\rp J^2\rb dk.
\eq

In order to obtain an analytical dispersion relation \'a la Rayleigh-Ritz, we use the following trial function for J - 

\beq
J = e^{-\mu\kt/2}
\eq

This trial function corresponds to a shear-layer type magnetic-field profile with field reversal as is appropriate for the present situation. This trial function also corresponds to a spatial structure that is in full agreement with the numerically calculated eigenfunction for this problem by Terasawa\cite{Ter}.

Using (32), (31) becomes

\beq
S = \sqrt{\pi} \lb \lp \ve +\frac{\mut}{\om}\rp \munh - 2\om\muh\rb + \sqrt{\pi}\frac{\st\mut}{\ve}\lb\munh-\sqrt{\frac{\pi \om}{\ve}}e^{\frac{\large\mu\om}{\large\ve}}\lel 1 -~{erf}\lp\sqrt{\frac{\mu\om}{\ve}}\rp\rel\rb.
\eq

In order to facilitate further devlopment with (33), it becomes necessary to simplify (33) by restricting the trial-function parameter $\mu$ to values such that $\lp\mu\om/\ve\rp\gg1$. (33) then becomes

\beq
S=\sqrt{\pi}\lb\lp\ve+\frac{\mut}{\om}\rp\munh-2\om\muh+\frac{\st}{2\om}\muh\rb.
\eq

The parameters $\mu$ and $\om$ may then be determined by imposing the conditions - 

\begin{subequations}
\beq
S=0
\eq
\beq
\frac{\pa S}{\pa\mu}=0
\eq
\end{subequations}

\noindent
which lead to

\beq
2\mut + (\st - 4\om^2)\mu+ 2\om\ve=0
\eq

\beq
6\mut + (\st - 4\om^2)\mu - 2\om\ve = 0.
\eq

Eliminating $\mu$, equations (36) and (37) lead to the dispersion relation - 

\beq
\om^2 - \sqrt{\ve\om} - \frac{\st}{4} = 0.
\eq

Rewriting (38) as

\beq
\lp \om + \frac{\sigma}{2}\rp \lp\om-\frac{\sigma}{2}\rp = \sqrt{\ve\om}
\eq

\noindent
one observes that the stable and unstable Hall branches become coupled in the presence of resistivity $(\ve\neq 0).$\footnote{\large It may be noted that resistivity does similar coupling of the electron-inertia and electron-compressibility branches of the collisionless tearing mode in a low-$\beta$ plasma (Shivamoggi \cite{Shiv}).}

According to (38), the resistive branch is given by the FKR result - 

\beq
\omega\sim\ve^{1/3}
\eq
\noindent
while the Hall branch is given by

\beq
\omega\sim\sigma.
\eq

On recalling that the Hall current is simply the effect due to ion inertia, one observes that (41) is apparently consistent with the previously found growth rate of the electron-inertia driven tearing mode $\om\sim d_e$ (Coppi \cite{Cop}).

Further, according to (40) and (41), the transition from the resisitive to the Hall branch occurs for 

\beq
\sigma\sim \ve^{1/3}
\eq

\noindent
in agreement with the scaling relation for the transition regime numerically established by Fitzpatrick \cite{Fit} in a driven Hall reconnection situation. 

For a weak Hall effect, (38) gives

\beq
\om\approx\ve^{1/3} + \frac{1}{6} \st \ve^{1/6}
\eq

\noindent
which shows that the Hall effect enhances the tearing mode growth rate in agreement with the numerically calculated result of Terasawa \cite{Ter}.

\indent On the other hand, for a weak resistive effect, (38) gives
\beq
\om\approx\frac {\sigma}{2} + \sqrt\frac{\ve}{2\sigma} 
\eq
The $\ve^{1/2}$(or   $\eta^{1/2}$) dependence of the correction to the growth rate in the Hall-effect dominated regime, indicated by (44) may be of some interest (Hassam [14] reported the same resistivity dependence for the total growth rate, however).
\section{Discussion}

In recognition of the difficulties resulting from the {\it triple-deck} structure associated with the Hall resistive tearing mode a variational method is used to formulate this problem. This method turns out to be very well suited for this purpose and leads to a convenient analytical dispersion relation for the Hall resistive tearing mode. This analytical dispersion relation -

\begin{itemize}
  \item recovers the FKR \cite{Fur} result for the resistive branch;
  \item gives a growth rate for the Hall branch which appears to be  consistent with the growth rate of the electron-inertia driven tearing mode given previously (Coppi \cite{Cop});
  \item recovers the sealing relation for the transition from the resistive regime to the Hall regime numerically established by Fitzpatrick \cite{Fit} in a driven Hall resistive reconnection situation.
 \end{itemize}

\section{Acknowledgments}

I acknowledge with gratitude helpful discussions with Drs. Luis Chacon and Mike Johnson.

\end{document}